\documentclass[12pt]{article}

\usepackage{latexsym}
\usepackage{amsthm}
\usepackage{amsmath}
\usepackage[dvips]{graphicx}
\usepackage{amssymb}

\newcommand{\be}{\begin{eqnarray}}
\newcommand{\ee}{\end{eqnarray}}

\begin{document}

\baselineskip=18pt

\setcounter{footnote}{0}
\setcounter{figure}{0} \setcounter{table}{0}

\begin{titlepage}

\begin{center}
\vspace{1cm}

{\Large \bf  On Tree Amplitudes in Gauge Theory and Gravity}

\vspace{0.8cm}

{\bf Nima Arkani-Hamed$^1$, Jared Kaplan$^{a,b}$}

\vspace{.5cm}

{\it $^{1,a}$ School of Natural Sciences, Institute for Advanced Study \\
Olden Lane, Princeton, NJ 08540, USA}

{\it $^b$ Jefferson Laboratory of Physics, Harvard University,\\
Cambridge, Massachusetts 02138, USA}

\end{center}
\vspace{1cm}

\begin{abstract}

The BCFW recursion relations provide a powerful way to compute tree
amplitudes in gauge theories and gravity, but only hold if some
amplitudes vanish when two of the momenta are taken to infinity in a
particular complex direction. This is a very surprising property,
since individual Feynman diagrams all diverge at infinite momentum.
In this paper we give a simple physical understanding of amplitudes
in this limit, which corresponds to a hard particle with (complex)
light-like momentum moving in a soft background, and can be
conveniently studied using the background field method exploiting
background light-cone gauge. An important role is played by enhanced
spin symmetries at infinite momentum--a single copy of a ``Lorentz"
group for gauge theory and two copies for gravity--which together
with Ward identities give a systematic expansion for amplitudes at
large momentum. We use this to study tree amplitudes in a wide
variety of theories, and in particular demonstrate that certain pure
gauge and gravity amplitudes do vanish at infinity. Thus the BCFW
recursion relations can be used to compute completely general gluon
and graviton tree amplitudes in any number of dimensions. We briefly
comment on the implications of these results for computing massive
4D amplitudes by KK reduction, as well understanding the unexpected
cancelations that have recently been found in loop-level gravity
amplitudes.

\end{abstract}

\bigskip
\bigskip

\end{titlepage}

\section{Introduction}

The textbook formulation of perturbative QFT as an expansion in
Feynman diagrams includes an enormous amount of unphysical off-shell
structure. This is particularly true of Yang-Mills theories and
General Relativity, where the gauge and diffeomorphism redundancies
are introduced to make Lorentz invariance and locality manifest.
While Lorentz invariance is very likely an exact property of Nature,
non-perturbative gravity makes it impossible to define off-shell
local observables, and therefore locality is a more suspect notion
at a fundamental level. It would therefore be interesting to find a
different formulation of QFT not relying so heavily on manifest
locality. Such a formulation might allow for a more natural
inclusion of gravity, much like the non-manifestly deterministic
least action formulation of classical mechanics generalizes more
naturally to quantum mechanics than Newton's laws. There is a more
down-to-earth reason for suspecting that another formulation of QFT
exists: on-shell gauge and gravity amplitudes, particularly for many
external legs, receive contributions from a huge number of Feynman
diagrams, but extensive cancelations take place and the final
results are strikingly simple, exhibiting regularities that are
invisible in the diagrammatic expansion \cite{MHV,BG,reviews}. The
simplicity of the final answer suggests that there should be another
way of computing the amplitudes more directly.

\subsection{BCFW Redux}

For tree amplitudes, a huge step in this direction was taken by
Britto, Cachazo, Feng \cite{BCF} and further clarified with Witten
\cite{BCFW}. Their work was an outgrowth of Witten's twistor
formulation of four-dimensional Yang-Mills \cite{Twistor,CSW}, which
is crucially tied to 4D physics. Indeed, the great simplicity of
maximal helicity violating amplitudes is due to their close
connection to self-dual solutions of the Yang-Mills equations of
motion \cite{Bardeen, Selivanov}, which is very special to four
dimensions. But the BCFW ideas do not rely on twistors or the
spinor-helicity formalism, and are instead a general property of QFT
in any number of dimensions, as we now review.

Consider the $n$-point amplitude $M(p_i,h_i)$ for massless particles
with $h_i$ ``helicities" in a general number $D$ of spacetime
dimensions. When we consider gauge theory, we will define
$M(p_i,h_i)$ such that the color factors are already stripped away.
We will also suppress the trivial overall multiplicative coupling
constant dependence. The key idea is to pick two external momenta
$p_j$,$p_k$, and to analytically continue these momenta keeping them
on-shell and maintaining momentum conservation. Specifically, BCFW
take \be p_{j} \to p_j(z) = p_{j} + q z \ \ \ \mathrm{and} \ \ \ p_k
\to p_k(z) = p_k - q z \ee where we must have $q \cdot p_{j,k} = 0$
, $q^2 = 0$. This is impossible for real $q$, but possible for
complex $q$. To be explicit, choose a Lorentz frame where $p_j,p_k$
are back to back with equal energy and use units where that energy
is $1$. Then, we can choose \be p_j  = (1,1,0,0;0..,0), \ p_k =
(1,-1,0,0;0,..0), \ q = (0,0,1,i;0,..0) \ee 
% Alternatively, we can
% keep all the momenta real but imagine that we are working in
% $SO(D-2,2)$ signature. 
Note that this deformation only makes sense
for $D \ge 4$.

What about the polarization tensors? Note that for gauge theory in a
covariant gauge,  $q = \epsilon^-_1 = \epsilon^+_2$. This makes it
natural to use a $+,-,T$ basis for spin 1 polarization vectors where
\be \epsilon_j^- = \epsilon_k^+ = q, \ \ \ \epsilon_j^+ =
\epsilon_k^- = q^*, \ \ \ \epsilon_T = (0,0,0,0,...,1,...,0) \ee
with $D-4$ different $\epsilon_T$ forming a basis in the transverse
directions. When the momenta are deformed, the polarization vectors
must also change to stay orthogonal to their associated momenta and
maintain their inner products. This requires \be \epsilon_j^-(z) =
\epsilon_k^+(z) = q, \ \ \epsilon_j^+(z) = q^* - z p_k, \ \
\epsilon_k^-(z) = q^* + z p_j, \ \ \epsilon_T(z) =
(0,0,0,0,...,1,...,0) \ee 
Alternatively, we can keep all the momenta and polarization vectors 
real but imagine that we are working in $SO(D-2,2)$ signature; 
this point of view will allow us to avoid subtleties when we take 
the complex conjugates of field derivatives.
Graviton polarization tensors are simply
symmetric, traceless products of these gauge polarization vectors. A
general product of polariation tensors including the antisymmetric
and trace parts gives amplitudes including a dilaton and
antisymmetric two index tensor field.

With this deformation, $M(p_i,h_i) \to M(z)$ becomes a function of
$z$. At tree level, $M(z)$ has an extremely simple analytic
structure -- it only has simple poles.  This follows from a
straightforward consideration of Feynman diagrams, as all
singularities come from propagators, which are simply \be
\frac{1}{P_J(z)^2} = \frac{1}{\left(\sum_{i \in J} p_i \right)^2}
\ee where $J$ is some subset of the $n$ momenta. Since $p_j(z) +
p_k(z)$ is independent of $z$, this only has non-trivial $z$
dependence when only one of $p_j(z)$ or $p_k(z)$ are included in
$J$.  Without loss of generality we take $j \in J$, in which case we
have \be \frac{1}{P_J(0)^2 - 2 z q \cdot P_J} . \ee This shows that
all singularities are simple poles located at $z_J = P_J(0)^2 / (2 q
\cdot P_J)$. Furthermore, the residue at these poles has a very
simple interpretation as a product of lower amplitudes: \be
\mathrm{res} M(z \to z_J) = \sum_h M(i \in J, p_i(z_J), h_i;
-P_J(z_J), h) \times M(i \notin J, p_i(z_J), h_i; P_J(z_J), -h) .
\ee where we have a sum over helicities for the usual reason,
guaranteed by unitarity, that the numerator of the propagator can be
replaced by the polarization sum on shell.

So far everything has been kinematical and true for an arbitrary
theory. What is remarkable is that for certain amplitudes in some
theories, $M(z \to \infty)$ vanishes. Now, meromorphic functions
that vanish at infinity are completely characterized by their poles;
if $M(z \to \infty) = 0$, we have $0 = \int dz/z M(z) = M(0) +
\mathrm{residues}$, and this gives us the BCFW recursion relation
\be M(0) = \sum_{J,h} M(i \in J, p_i(z_J); h_i, -P_J(z_J), h)
\frac{1}{P_J^2} M(i \notin J, p_i(z_J), h_i; P_J(z_J), -h) \ee where
$h$ indicates a possible internal helicity. The lower amplitudes are
on-shell (in complexified momentum space), because all the momenta
are on shell though evaluated at a complex $z=z_J$. These recursion
relations produce a higher-point amplitude by sewing together
lower-point on shell amplitudes.

\begin{figure}[h]
\begin{center} \label{FigBCFRR}
\includegraphics[width=17cm]{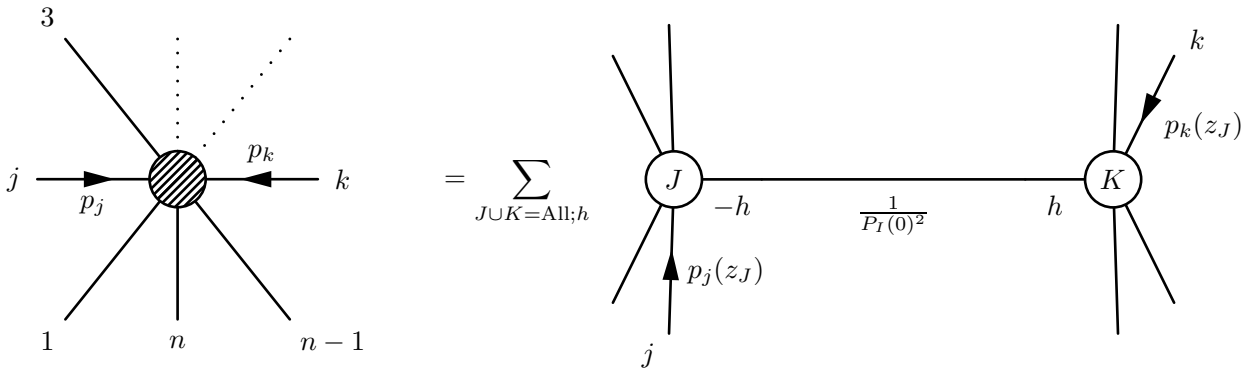}
\caption{\small{The BCFW recursion relation computes an $n$-point
amplitude by sewing together lower-point amplitudes with (complex)
on-shell momenta.}}
\end{center}
\end{figure}

Of course the strategy of determining amplitudes directly from their
singularities is a familiar and central theme of the S-matrix
program. However, the old ideas were largely restricted to $2 \to 2$
scattering and the complexification of the Mandelstam $s,t,u$
variables, and the generalization to higher-point amplitudes was not
clear. Over the past twenty years, S-matrix ideas have had a
resurgence, as it has become increasingly clear that they provide
powerful methods for computing field theory amplitudes, for instance
as in  the unitarity methods of Bern, Dixon, Dunbar and Kosower
\cite{BD}. The BCFW recursion relations are another step in this
direction. Indeed, the BCFW deformation of momenta can be viewed as
a correct general procedure for complexifying on-shell momenta and,
at least at tree level, the recursion relations beautifully fulfil
the S-matrix dream of dealing directly with on-shell amplitudes
without reference to an off-shell Lagrangian.

\subsection{Surprising Behavior of $M(z)$ as $z \to \infty$}

In order to derive these recursion relations, it was necessary to
assume $M(z \to \infty)$ vanishes. But this is far from obvious, and
is not even true for every amplitude in a general theory. Indeed,
naively it is never true! Consider for instance $\phi^4$ theory,
here the $ 2 \to 2$ amplitude is momentum independent and hence $z$
independent. With more external lines there are propagators that as
we have seen above fall as $1/z$, however, there is always a diagram
with the $\phi^4$ interaction involving $j,k$ and two other lines,
with each of the lines separately attaching to two separate sets of
external states. The large momentum does not flow through any of the
propagators, so general amplitudes go to a constant as $z \to
\infty$ \be M^{\phi^4}(z) \to z^0 \ee

\begin{figure}[h]
\begin{center} \label{FigDangerousDiagram}
\includegraphics[width=15cm]{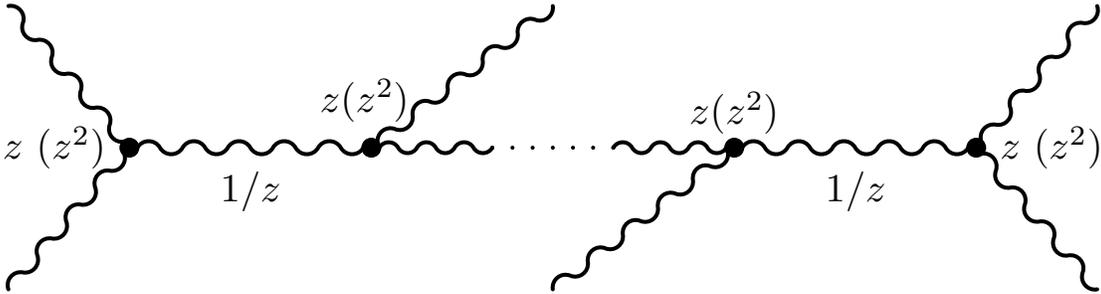}
\caption{\small{Contributions to the analytically continued
amplitudes $M(z)$ from individual Feynman diagrams diverge as $z \to
\infty$ for gauge theories and gravity. This is due to the vertices
that grow as $z,z^2$ for gauge theory and gravity, which
overcompensate for the $1/z$ scaling of propagators.}}
\end{center}
\end{figure}

The situation seems even worse with gauge theory and gravity, where
there are momentum dependent vertices that scale as $z$ in gauge
theory and $z^2$ in gravity, and so we might expect that $M(z)$
diverges as $z \to \infty$. Indeed, naively in gauge theory \be
M_{\mathrm{naive}}^{-+}(z) \to z, \ \ M_{\mathrm{naive}}^{--/++}(z)
\to z^2, \ \ M_{\mathrm{naive}}^{+-}(z) \to z^3, \ee where the $+$
and $-$ signs represent the different gauge boson helicity states,
and different powers of $z$ result from the growth of some of these
polarization vectors as $z \to \infty$. In gravity the naive
divergence grows as power of the number of external legs, for
example \be M_{\mathrm{naive}}^{--,++}(z)  \to z^{n - 1} \ \ \
\cdots, \ \ M_{\mathrm{naive}}^{++,--}(z) \to z^{n+3} . \ee

Nonetheless, at least for some $h_j,h_k$, the amplitudes \emph{do}
vanish as $z \to \infty$!  Even when they do not, they often diverge
more mildly than the naive expectation; this neatly encapsulates the
heavy cancellations that take place in the explicit evaluation of
Feynman digrams. For instance for 4D gauge theories, BCFW showed
that \be M^{-+}(z), \ M^{--}(z), \ M^{++}(z) \to \frac{1}{z} \ \ \
\mathrm{and} \ \ \ M^{+-} \to z^3. \ee BCFW gave a simple
diagrammatic proof for $(-+)$,  whereas the $(--)$ and $(++)$ cases
needed a different argument based on MHV diagrams and the CSW
recursion relations, which are very special to $D=4$. Note that for
the recursion relations to hold, it is not necessary for $M(z)$ to
vanish for all $h_{jk}$ helicities. For gauge theory it suffices to
have e.g. $M_{- h}(z \to \infty)$ all vanish, since for any
amplitude, we can always choose $q$ to co-incide with
$\epsilon_1^-$; for gravity it suffices for $M_{-- h}(z \to \infty)$
to vanish for the same reason.

In 4D gravity, surprisingly good behavior for amplitudes was first
observed by \cite{grav1}, \cite{BjerrumBohr1}. Cachazo et. al. then showed in beautiful
papers \cite{grav3,grav4} that \be M^{--,++} \to 1/z^2 \ \ \
\mathrm{and} \ \ \ M^{++,--} \to z^6 \ee Their analysis for the
$(--,++)$ case involved intricate diagrammatic and combinatorial
recursion arguments, and the $(--,--)$, $(++,++)$ and $(++,--)$
cases were only controlled for MHV amplitudes. Subsequently, Bern
et. al. showed that the $(++,--)$ scaling holds in general, and
found the general scaling for all helicity combinations up to 10
external legs \cite{Bern3}, \cite{BjerrumBohr2}. Note that the $z$ scaling conforms to
the famous KLT pattern $M_{\mathrm{grav}} \sim M_{\mathrm{gauge}}
\times M_{\mathrm{gauge}}$ \cite{KLT}, (though KLT only controls
these amplitudes for $2 \to 2$ scattering while for more legs, term
by term the amplitudes are nearly as divergent as with standard
Feynman diagrams). It is remarkable that, far from being
uncontrollably divergent, certain gravity amplitudes are even better
behaved at infinity than their gauge counterparts, which are in turn
better behaved than the simplest scalar field theories!

There is clearly simplicity and a pattern to $M(z \to \infty)$ in
gauge theory and gravity. What is known so far in generality is
restricted to four dimensions, and the techniques used to understand
the large $z$ behavior differ from case to case and do not
illuminate this pattern, or tell us what to expect for general
theories in any number of dimensions. In this paper, we will develop
a more transparent, physical understanding of the large $z$ behavior
that is valid in any number of dimensions; apart from a clearer
understanding of the physics this also immediately generalizes the
BCFW recursion relations to gauge and gravity amplitudes in any
number of dimensions.

\section{Understanding $M(z \to \infty)$ in YM and GR}

We begin by observing that as $z \to \infty$, the momenta $p_{j,k}$
tend to $\pm z q$, and if we think of one as ingoing and the other
as outgoing, this is simply a limit where a hard light-like particle
is shooting through a soft background. For {\it real} momenta, this
is the very familiar eikonal limit -- indeed the soft collinear
effective theory \cite{SCET} provides a natural formalism for
studying physics in this regime, though we won't make any use of the
machinery of this subject in our analysis. Intuitively, a highly
boosted particle will not be ``much" scattered by the background,
and its helicity should be conserved. This is not precisely our
situation because our hard light-like momentum is complex (or
equivalently real but in $(D-2,2)$ signature). We will see that we
can understand the behavior of the amplitudes at large $z$ as an
expansion in $1/z$, that both quantifies the intuition for real
momenta and can be used to understand the scalings for the complex
momenta of interest. Since we only care about the $z$ dependence of
the amplitudes, all the soft physics can be absorbed into
determining some classical background, and the single hard line can
be studied by looking at quadratic fluctuations about this
background. Another natural approach would be to use the worldline
formalism for a particle propagating in a background field
\cite{WL}, but we will see that the standard field theory techniques
are already very simple.

We proceed to study the large $z$ behavior of amplitudes in theories
of increasing complexity; scalar QED, scalar Yang-Mills, Yang-Mills,
gravity coupled to a scalar, to a photon, and finally gravity
itself.

\subsection{Scalar QED Amplitudes}

We start with scalar QED as a simple warm-up. We will dwell on a
number of issues at greater length in this subsection where they can
be explored in the simplest setting, and abbreviate the analagous
discussion in the subsequent subsections.

Consider amplitudes $M_n$ with exactly two external scalar lines and
$n$ external photons. We choose to analytically continue the momenta
corresponding to the scalar lines. The scalar Lagrangian is \be L =
D_\mu \phi^* D^\mu \phi . \ee where we view $A_\mu$ as a background
field in which the highly boosted scalar particle propagates. Note
that naively $M(z) \to z$ for large $z$, since the
scalar-scalar-photon vertex has a momentum that scales as $z$. We
will see that however that $M_n(z)$ is in fact much better behaved.

In considering the amplitude for large momentum, we immediately run
into the problem that `large' momenta for $\phi$ is not a
gauge-invariant.  The natural way to deal with this
issue is to perform a field re-definition (as in SCET \cite{SCETforQED}), 
stripping off a Wilson line from the field \be \phi(x) = W_n(x) \tilde{\phi}(x), \ \
\mathrm{where}  \ \ \ W_n(x) = \exp \left( i \int_{-\infty}^0
d\lambda \ n_\mu A^\mu(x + \lambda n) \right) \ee and $n^\mu$
determines the direction of the Wilson line stretching from the
point $-\infty$ to $x$.  Since $\tilde{\phi}(x)$ is gauge invariant,
its ordinary derivative gives a gauge invariant definition of
momentum. The Lagrangian becomes \be L =  (W_n
\partial_\mu \tilde{\phi} + D_\mu W_n \tilde{\phi})^* (W_n
\partial^\mu \tilde{\phi} + D^\mu W_n \tilde{\phi}) . \ee and the only
terms that grow as $z \to \infty$ are the cross terms such as \be
\partial_\mu \tilde{\phi}^* \tilde{\phi} W_n^* D^\mu W_n
\to i z q_\mu \tilde{\phi}^* \tilde{\phi}  W_n^* D^\mu W_n . \ee
Now, $W_n^* D_\mu W_n$ is gauge invariant and a trivial computation
shows \be q_\mu W_n^* D^\mu W_n = i \int_{-\infty}^0 d \lambda \,
q_\mu F^{\mu \nu}(x + n \lambda) n_\nu \ee Choosing $n_\mu = q_\mu$,
corresponding to the Wilson line pointing in the light-like
direction of the large momentum $q_\mu$ itself, this combination
vanishes due to the antisymmetry of $F^{\mu \nu}$, showing in a
gauge-invariant way that there are no physical $O(z)$ vertices in
this theory.

Of course, there is a much simpler way of eliminating gauge
redundancy and working with gauge invariant quantities--we can
simply choose a gauge! The only $O(z)$ interactions from the
original lagrangian involve $q \cdot A$ terms, so the natural choice
is $q$-light cone gauge, with \be q \cdot A = A_- = 0. \ee Indeed,
the gauge invariant $-i W_q^* D_\mu W_q$ we encountered in the
previous paragraph is nothing but $A_\mu$ itself in this light-cone
gauge.

The utility of $q$-light cone gauge for gauge theory computations
was recognized by Chalmers and Siegel in \cite{ChS}, where it was
dubbed ``space-cone" gauge. Vaman and Yao \cite{VY} made use of this
gauge to give an understanding of the BCFW rules for gauge theory.
In our discussion of Yang-Mills theory and especially gravity in the
next subsections, we will simply go to light-cone gauge rather than
give the analog of the explicitly gauge invariant description in
terms of Wilson lines as in the above.

We have seen that there are no interaction vertices at $O(z)$. The
scalar propagator is proportional to $1/z$, so all diagrams with at
least one scalar propagator vanish as $z \to \infty$. Of course, due
to the four-point interaction vertex, $M_2(z)$ goes to a constant at
large $z$, however, all $n$-point amplitudes with $n>2$ photons must
have at least one scalar propagator, and so we conclude \be M_2(z)
\to z^0  \ \ \ \mathrm{and} \ \ \ M_{n>2}(z) \to \frac{1}{z}, \ee In
fact clearly, for large $n$ there are more propagators and the
amplitudes are suppressed by higher powers of $z$. Thus the BCFW
Recursion Relations apply to scalar QED with two external scalars
and at least three external photons.

It is worth noting that the better behavior of the amplitudes at
large $z$ is not merely a consequence of gauge invariance.  The
addition of higher dimension gauge invariant operators, for example
the operator $D_\mu \phi^* D_\nu \phi F^\mu_\alpha F^{\nu \alpha}$,
leads to vertices with positive powers of $z$. The good behavior of
$M(z \to \infty)$ is thus a special feature of the two-derivative
Lagrangian neglecting any higher-dimension operators.

\subsection{Dominant Large $z$ Behavior}

Our arguments above for eliminating the $O(z)$ vertices either using
Wilson lines or going to light-cone gauge were a little too quick;
there is a subtlety that is irrelevant for scalar QED but will be
important in the rest of the examples, and will allow us to isolate
the dominant large $z$ amplitude in all cases. Consider again our
argument that $q_\mu W_n^* D^\mu W_n$ vanishes for $n_\mu = q_\mu$.
It is true that the integrand $q_\mu F^{\mu \nu}(x + n \lambda)
n_\nu$ vanishes as $n_ \mu \to q_\mu$, but there is also a
semi-infinite integral over $\lambda$, so one might have a $0 \times
\infty$ ambiguity. Indeed, simply going to momentum space we can
perform the $\lambda$ integral and find \be \left[q_\mu W_n^* D^\mu
W_n\right](p) = \frac{q_\mu F^{\mu \nu}(p) n_\nu}{p \cdot n} =
\frac{(A(p) \cdot q)(p \cdot n) - (A(p) \cdot n)(p \cdot q)}{p \cdot
n} \ee As long as $p \cdot q \neq 0$, we can take $n_\mu \to q_\mu$
and this vanishes as expected. But if $p \cdot q = 0$, then we have
a problem -- as $n \to q$ there is a $0/0$ cancelation and we find
\be {\rm lim}_{n \to q} \left[q_\mu W_n^* D^\mu W_n\right](p) = A(p)
\cdot q \ee goes to a constant.

This subtlety also shows up in going to light-cone gauge, and indeed
obstructs making this gauge choice. In order to choose light cone
gauge, in momentum space we need to find a $\Lambda(p)$ so that \be
q^\mu A_\mu(p) + i q^\mu p_\mu \Lambda(p) = 0, \ee but this is
impossible if $p \cdot q = 0$ unless the gauge transformation
becomes singular $\Lambda(p) \to \infty$.

We conclude that if there is a component of the background field
carrying a a momentum $p$ such that $p \cdot q = 0$, there {\it is}
a physical $O(z)$ vertex. In scalar QED, the photons are
non-interacting so the background field is a sum of the external
plane waves, and the background field momenta are just the external
momenta. In non-Abelian theories there are self-interactions, and
the possible components of background field momenta are simply sums
of subsets of the external soft particle momenta. Thus generic
momenta will indeed have $p \cdot q \neq 0$ and the subtlety above
is irrelevant.
\begin{figure}[h]
\begin{center} \label{FigSubtlety}
\includegraphics[width=15cm]{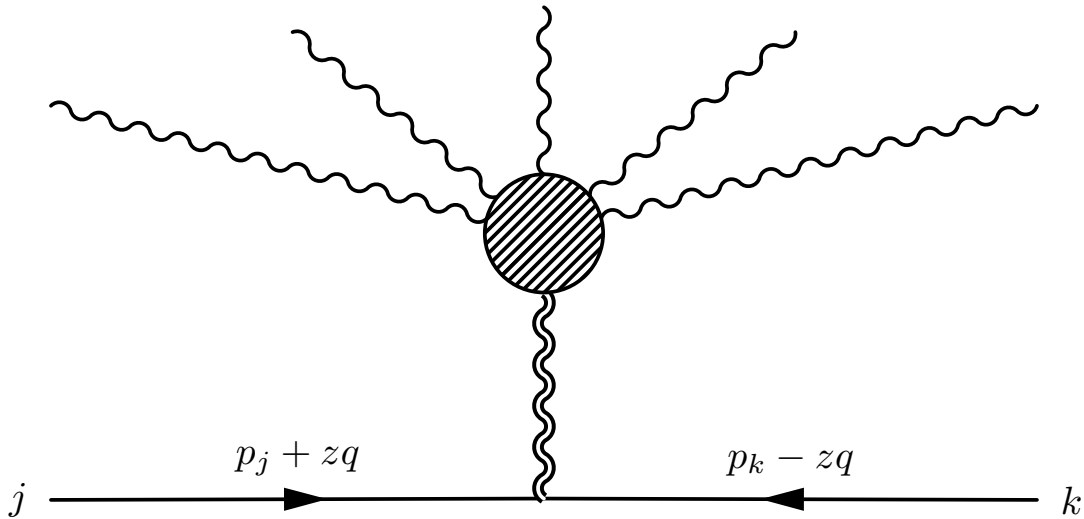}
\caption{\small{The unique set of diagrams, for which light-cone
gauge is singular, and which dominate large $z$ amplitudes.}}
\end{center}
\end{figure}
But the sum over {\it all} the external momenta must equal $-(p_j +
p_k)$, which is orthogonal to $q$. Therefore, there is a unique set
of diagrams where this subtlety is relevant -- those diagrams where
the two external analytically continued lines meet in a three point
vertex with a single background field, which then connects to {\it
all} remaining external fields.

In scalar QED this never occurs, because the only such diagram would
only include one photon. However, this unique class of diagrams do
occur and will be of importance in non-Abelian gauge theory and even
more so in gravity. The $z$ scaling of these diagrams is the naive
one corresponding to the number of momenta in the vertex -- up to
$O(z)$ for gauge theory and $O(z^2)$ for gravity, and therefore
these unique diagrams dominate the amplitude at large $z$.

We can rephrase this discussion directly in the $q$-lightcone gauge.
Here, the only singularities occur when we eliminate the ``+"
components of the gauge or gravity fields, for instance in gauge
theory we have $\partial_- A_+ - \partial_i A^i = 0$, so that
$A_+(p)$ has a $1/p_-$ singularity. As we have seen, there is a
unique set of diagrams where  $p_-$ vanishes for an internal
background line. Since none of the other diagrams have this
singularity and the full amplitude is gauge invariant, we conclude
that the $1/p_-$ factors are always cancelled by $p_-$ factors in
vertices to leave a non-singular result from this unique class of
diagrams alone. But, since the light-cone gauge choice is
non-singular for the other diagrams where the leading $z$ dependence
is eliminated, we can also conclude that nothing can cancel the
naive $z$ scaling of our unique set of diagrams either. Thus we have
identified the leading contributions to the amplitude at large $z$.

\subsection{Scalar-Yang-Mills Amplitudes}

We move on to consider scalar Yang-Mills, where the above discussion
can be seen in action in the simplest setting. In this case, the
Lagrangian is \be L = {\rm tr} D_\mu \phi^* D^\mu \phi \ee The soft
background field $A_\mu$ is the solution of the Yang-Mills equations
of motion that non-linearly completes the sum over the plane waves
corresponding to the soft external gluons (formally, to insure that
only connected diagrams are summed, the soft polarization vectors
should be thought of as being anti-commuting, see e.g.
\cite{Selivanov}). We will be assuming that $\phi$ is in the adjoint
representation of the gauge group, since this will allow greater
cohesion when we move on to pure Yang-Mills in the next section.

Now for the physics. We can again fix $q$-lightcone gauge to
eliminate the $O(z)$ vertices as in scalar QED, but now, due to
gluon self-interactions, we have the unique set of diagrams
described above, with an arbitrary number of external gluons that
only include a single scalar-gluon vertex, where two scalars with
momenta $p_j(z)$ and $p_k(z)$ meet an off-shell gluon field with
momentum $-(p_j + p_k)$. Since the two external scalar lines in
these diagrams couple directly through a 3-point vertex, the scalars
must be adjacent in color at leading order in $N_c$ (planar
diagrams). As we argued in the previous subsection, these diagrams
scale as $O(z)$.

For scalars that are not adjacent in color, the self-interactions of
the gluons allow diagrams without scalar propagators to contribute,
so the amplitude will be $O(1)$.  So we have found \be M_{j k \,
\mathrm{adj. \, in \, color}}(z) \to z \ \ \ \mathrm{and} \ \ \ M_{j
k \, \mathrm{non-adj. \, in \, color}}(z) \to z^0 \ee This behavior
can be readliy verified in simple examples where explicit amplitudes
are known, such as $2 \to 2$ scattering. Thus unlike scalar QED,
there are no BCFW Recursion Relation in scalar Yang-Mills theory
when the external scalar lines are analytically continued.

\subsection{Gluon Amplitudes and Enhanced Spin Symmetry as $z \to \infty$}

For gluon amplitudes in gauge theory, we will study the two particle
amplitude $M^{\mu \nu}$ in a soft background field, where $\mu$ and
$\nu$ are the Lorentz indices that will be contracted with the
polarization vectors of the highly boosted particle (we suppress the
color indices on the amplitude). A new feature will be the presence
of an enhanced spin ``Lorentz" symmetry which will largely control
the large $z$ behavior of the amplitude. Together with a simple
application of the Ward identity, this will yield the desired large
$z$ scalings.

Expanding the gauge field ${\cal A}_\mu = A_\mu + a_\mu$ where
$A_\mu$ is the background and $a_\mu$ the fluctuation, the quadratic
Lagrangian for $a_\mu$ is \be L = -\frac{1}{4} \, \mathrm{tr} D_{[
\mu} a_{\nu ]} D^{[ \mu} a^{\nu ]} + \frac{i}{2} \mathrm{tr} \,
[a_\mu, a_\nu] F^{\mu \nu} \ee where $D$ is the $A$-covariant
derivative.  As usual in the background field method, we have two
types of gauge symmetry -- gauge transformations of $a$ and of $A$.
We fix the $a$ gauge freedom in the usual way by adding a gauge
fixing term $(D_\mu a^\mu)^2$ to the lagrangian. The gauge fixed
Lagrangian is \be \label{yangmillsL} L = -\frac{1}{4} \mathrm{tr}
D_{\mu} a_{\nu } D^{\mu} a^{\nu} + \frac{i}{2} \mathrm{tr} [a_\mu,
a_\nu] F^{\mu \nu} \ee.

Note that the first term in this Lagrangian is the only one with the
potentially $O(z)$ vertices, and hence dominates in the amplitude as
$z \to \infty$. But this first term also enjoys an enhanced ``spin"
symmetry -- a Lorentz transformation acting on the $\nu$ indices of
$a_\nu$ alone. We'll call this a spin ``Lorentz" invariance, since
the actual Lorentz invariance is explicitly broken by the
non-vanishing background field even for the first term in the
Lagrangian. To make this symmetry more explicit, we trivially
re-label indices so that the Lagrangian is \be L = -\frac{1}{4}
\mathrm{tr} \, \eta^{a b} D_{\mu} a_{a} D^{\mu} a_{b} + \frac{i}{2}
\mathrm{tr} [a_a, a_b] F^{a b} \ee As already noted, the first term
dominates the large $z$ amplitude but is spin ``Lorentz" invariant,
while the second term breaks the Lorentz symmetry as an
antisymmetric tensor. This allows us to determine the form of
$M^{ab}$. Since all the $O(z)$ vertices come from the first term,
and only repeated use of these vertices can possibly give an
amplitude that scales as $z$, the part of the amplitude that scales
as $z$ must also be proportional to $\eta^{a b}$. This reflects the
intuitively familiar fact that the helicity of a highly boosted
particle blasting through a soft background is conserved. The first
contribution that breaks the ``Lorentz" spin symmetry arises from a
single insertion of vertices coming from the second term in the
Lagrangian, and must be antisymmetric in $(ab)$, just as $F^{ab}$
is. Further insertions give more powers of $1/z$ which multiply
general matrices in $(ab)$ space. Thus, the ``Lorentz" symmetry
guarantees that the amplitude has the form \be M^{a b } = (c z +
\cdots) \eta^{a b } + A^{a b} + \frac{1}{z} B^{a b} + \cdots \ee
where $A^{a b}$ is antisymmetric in $(ab)$.

We can now find the $z$-dependence of the amplitude for various
helicity combinations by contracting our ansatz for $M^{ab}$ with
polarization vectors. The Ward identity further constrain $M^{ab}$.
For Yang-Mills theory, the Ward identity says that \be p_{ja}(z)
M^{ab} \epsilon_{k b} = 0 , \ee and similarly with $j$ and $k$
reversed, (but recall that $p_{ja} M^{ab} \neq 0$ when the second
$b$ index is not contracted with $\epsilon_k$). This implies \be
p_{j a}(z) M^{a b} \epsilon_{k b\nu} = 0 \ \ \ \implies \ \ \ q_a
M^{a b} \epsilon_{k b} = -\frac{1}{z} p_{1 a} M^{a b} \epsilon_{k b}
\ee which is extremely useful because $\epsilon_1^- = q$, so we can
use it to replace $\epsilon_j^- \to -\frac{1}{z} p_{j}$. Using this
information, let us look at the large $z$ amplitudes for a few
helicity combinations. Consider first $M^{-+}$; recall that this was
the only case that was understood by BCFW directly in terms of
Feynman diagrams \be M^{-+} & = & \epsilon_{j a}^- M^{a b}
\epsilon_{k b}^+ =  -\frac{1}{z} p_{j a} \left[(c z + \cdots)
\eta^{ab} + A^{ab} + \frac{1}{z} B^{ab} + \cdots \right] q_b \nonumber \\
&=& -\frac{1}{z} p_{j a} A^{ab} q_b + O(1/z^2) \to \frac{1}{z} \ee
as $z \to \infty$.  A more non-trivial case is \be M^{--}(z) & = &
\epsilon_{j a}^-  M^{a b} \epsilon_{k b}^- =
  -\frac{1}{z} p_{j a}
\left[(c z + \cdots) \eta^{ab} + A^{ab} + \frac{1}{z} B^{ab} +
\cdots \right]
(q_b^* + z p_{j b}) \nonumber \\
& = & -\frac{1}{z} p_{j a} A^{a b} q^*_b - \frac{1}{z} p_{j a} B^{a
b} p_{j b} +  O(1/z^2) \to \frac{1}{z} \ee as $z \to \infty$; note
that we have used the fact that $q^* \cdot p_j$, $p_j^2$, and $p_{j
a} p_{j b} A^{ab}$ all vanish, the last due to the anti-symmetry of
$A^{ab}$. As a last example before we simply list the results, let
us consider \be M^{+-}(z) = \epsilon_{j a}^+  M^{a b} \epsilon_{k
b}^- & = & (q_a^* - z p_{k a}) \left[(c z + \cdots) \eta^{ab} +
A^{ab} + \frac{1}{z} B^{ab} + \cdots \right]
(q_b^* + z p_{j b}) \nonumber \\
&\to& z^3. \ee The following table displays the general results for
gauge theory:
\begin{center}
\begin{tabular}{|c|c|c|c|}
\hline
$\epsilon_1 \backslash \epsilon_2$&$-$&$+$&$T$\\
\hline \hline
$-$&$1/z$&$1/z$&$1/z$\\
\hline
$+$&$z^3$&$1/z$&$z$\\
\hline
T1&$z$&$1/z$&$z$\\
\hline
T2&$z$&$1/z$&$1$\\
\hline
\end{tabular}
\end{center}
The difference between T2 and T1 is simply whether or not
$\epsilon_j^T \cdot \epsilon_k^T = 0$, respectively. We have checked
these results for the 2 to 2 amplitude in any number of dimensions
by using results in the literature. Since $M(z \to \infty)$ vanishes
for all the ($-,h$) helicity combinations, BCFW recursion relations
can be used to compute tree gluon amplitudes in any number of
dimensions.

Note that for this discussion, we did not use the $q$-lightcone
gauge for the background; exploiting this gauge gives us further
information. As before, $q$-lightcone gauge eliminates the $O(z)$
vertices except for the unique set of diagrams -- but as with the
scalar-YM case, these diagrams only exists if the hard momentum
lines are adjacent in color. This is also true of sub-leading
vertices coming from a single insertion of the $F^{ab}$ interaction.
Therefore, for non-adjacent colors, $M^{ab}$ begins at $O(1/z)$.
This is enough to guarantee that e.g. $M^{-+}$ scales as $1/z^2$ and
not $1/z$, and that $M^{+-}$ scales as $z^2$ rather than $z^3$, as
can again be confirmed for explicit $2 \to 2$ scattering amplitudes.
The $q$-lightcone gauge will play a more important role in
controlling the large $z$ amplitudes in gravity below.

\subsection{Scalar-Graviton Amplitudes}

Now we consider gravity, beginning with the case two scalar-graviton
amplitudes, where we analytically continue the scalar momenta. The
Lagrangian is \be L = \frac{1}{2} \sqrt{-g} g^{\mu \nu}
\partial_\mu \phi \partial_\nu \phi . \ee Naively, there are now
$z^2$ and $z$ vertices, and the amplitudes should blow up with
increasing powers of $z$ for more graviton legs.  However, using
diffeomorphism invariance we can choose light-cone gauge for the
background $g^{\mu \nu}$; taking $q_\mu$ to point in the  $+$
direction the gauge choice is \be g^{++} = g^{+i} = 0 \ \ \
\mathrm{and} \ \ \ g^{+-} = 1. \ee Note that equivalently, writing
$g_{\mu \nu} = \eta_{\mu \nu} + h_{\mu \nu}$, the gauge choice is
$h_{- \mu} = 0$.

Light-cone gauge eliminates the $O(z^2)$ vertices, but again we have
our unique diagram when the two external scalar lines couple to a
single insertion of the background $g^{\mu \nu}$ with an $O(z^2)$
vertex. Thus \be M(z) \to z^2. \ee This is much better behaved than
the naive expectations, reflecting the power of exploiting
background light-cone gauge for gravity amplitudes, although since
$M(z)$ still diverges there are no BCFW recursion relations using
analytically continued scalar momenta.

\subsection{Photon-Graviton Amplitudes}

We now move on to consider two photon-graviton amplitudes where we
analytically continue the photon momenta, corresponding to studying
a hard photon moving in a soft gravitational background. Our
experience with gauge theory suggests that we exploit an enhanced
spin symmetry at infinite momentum. To make this manifest, we need
to introduce Lorentz (rather than space-time) indices. For this
purpose, we use the vielbein $e_\mu^a$; this introduces, in addition
to the usual diffeomorphism redundancy, a gauge Lorentz redundancy
acting on the $a$ indices, with the associated connection
$\omega_\mu^{ab}$. The standard relation to the metric variables are
\be g_{\mu \nu} = e_{\mu a} e_{\nu b} \eta^{ab}, \ \ \ \omega_{\mu a
b} = e^\nu_a \nabla_\mu e_{\nu b}. \ee Since these fields connect
the asymptotic lorentz tensor structure to the local geometry, they
are exactly what we need. The Lorentz gauge redundancy is useful; in
addition to fixing the diffeomorphism redundancy for the metric by
choosing metric light-cone gauge, we will fix the extra Lorentz
redundancy by fixing light-cone gauge for $\omega_{\mu a b}$.

Let us now consider the Lagrangian for a photon in a gravitational
background \be L = -\frac{1}{4}\sqrt{-g} g^{\mu \alpha} g^{\nu
\beta} \nabla_{[ \mu} A_{\nu]} \nabla_{[ \alpha} A_{\beta]} . \ee If
we add the gauge fixing term $(\nabla_\mu A^\mu)^2$, we obtain \be L
= -\frac{1}{2}\sqrt{-g} g^{\mu \alpha} g^{\nu \beta} \nabla_\mu
A_\nu \nabla_\alpha A_\beta \ee where we have dropped a term
proportional to $R^{\mu \nu}$ because it vanishes on the background
field equations.  We introduce the vielbein so that \be A_\mu =
e^a_\mu A_a, \ \ \nabla_\nu A_\mu  = e^a_\mu D_\nu A_a, \
\mathrm{with}  \ \ D_\nu A_a =
\partial_\nu A_a + {\omega_{\nu a}}^c A_c \ee
The Lagrangian becomes \be L = - \sqrt{-g} g^{\mu \nu} \eta^{ab}
(\partial_\mu A_a + {\omega_{\mu a}}^c A_c) (\partial_\nu A_b +
{\omega_{\nu b}}^d A_d). \ee  Note that again, the two-derivative
interactions which dominate the amplitude at large $z$ respect a
spin ``Lorentz" invariance, broken by the subleading interactions
through non-vanishing $\omega_{ab \mu}$ which is antisymmetric in
$(ab)$. Choosing light-cone gauge for both $g^{\mu \nu}$ and
$\omega_{\mu a b}$, \be g^{++} = g^{+i} = 0, \, g^{+-} = 1 \ \
\mathrm{and} \, \, \omega^+_{ab} = 0   \ ,  \ee we see that there
are no $O(z^2)$ vertices and the only $O(z)$ vertices preserve the
spin Lorentz invariance; except for the by now familiar unique set
of diagrams. Thus $O(z^2)$ interactions must be proportional to
$\eta^{ab}$, while the $O(z)$ interactions not proportional to
$\eta^{ab}$ must involve a single insertion of the connection
$\omega_{\mu ab}$ which is anti-symmetric in $(ab)$ and so gives an
anti-symmetric contribution to $M^{ab}$. We therefore find \be
M^{ab} = c z^2 \eta^{ab} + z A^{ab} + B^{ab} + \cdots \ee where
$A^{ab}$, like $\omega$, must be antisymmetric, and $B^{ab}$ is
arbitrary. Using the Ward Identity as in the Yang-Mills section, the
large $z$ behavior of the amplitude for all helicity combinations
are given by
\begin{center}
\begin{tabular}{|c|c|c|c|}
\hline
$\epsilon_1 \backslash \epsilon_2$&$-$&$+$&$T$\\
\hline \hline
$-$&$1$&$1$&$1$\\
\hline
$+$&$z^4$&$1$&$z^2$\\
\hline
T1&$z^2$&$1$&$z^2$\\
\hline
T2&$z^2$&$1$&$z$\\
\hline
\end{tabular}
\end{center}
We have checked that these results agree with the known amplitudes
for the $2 \to 2$ graviton-photon scattering amplitudes -- a
non-trivial check, since individual Feynman diagrams diverge at
least as fast as $z^2$, with possible additional $z$'s coming from
contraction with the polarization vectors.

\subsection{Amplitudes in General Relativity}

We finally apply the lessons above to prove the BCFW Recursion
Relations for graviton amplitudes. Of course the results can be
anticipated via the KLT relations which express graviton amplitudes
as products of Yang-Mills amplitudes $M_{{\mathrm grav}} \sim
M_{{\mathrm gauge}} \times M_{{\mathrm gauge}}$. Indeed we will use
a simple and natural trick \cite{dilaton} that was originally
developed to help make the KLT relations manifest in GR, in order to
manifest an even larger spin ``Lorentz" invariance for graviton
amplitudes, which will determine the large $z$ behavior we seek.

The quadratic lagrangian for a gravitational fluctuation $h_{\mu
\nu}$ about an arbitrary background $g_{\mu \nu}$ is, after adding
the standard background de-Donder gauge fixing term \cite{gravback}
\be L & = & \sqrt{-g} \left[\frac{1}{4} g^{\mu \nu} \nabla_\mu
h_\alpha^\beta \nabla_\nu h^\alpha_\beta - \frac{1}{8} g^{\mu \nu}
\nabla_\mu h_\alpha^\alpha \nabla_\nu h_\beta^\beta - h_{\alpha
\beta} h_{\mu \nu} \frac{1}{2} R^{\beta \mu \alpha \nu} +
\frac{1}{2} g^{\mu \nu}
\partial_\mu \phi
\partial_\nu \phi \right]  \ee where we have used to the background
field equations to set $R_{\mu \nu} = 0$, and we have included an
extra dilaton field $\phi$.

Since dilaton number is conserved, at tree level the dilaton will
decouple from amplitudes involving only spin-2 graviton external
states. We have included it anyway because it will allow us to
perform a field re-definition that eliminates the
$h_{\alpha}^\alpha$ terms in the Lagrangian, which will enable us to
manifest two separate copies of a spin ``Lorentz" invariance acting
separately on the left and right indices of $h$. The field
redefinition is \be h_{\mu \nu} \to h_{\mu \nu} + g_{\mu \nu}
\sqrt{\frac{2}{D-2}} \phi, \ \ \ \ \ \ \ \ \ \ \phi \to \frac{1}{2}
g^{\mu \nu} h_{\mu \nu} + \sqrt{\frac{D-2}{2}} \phi \ee so that the
lagrangian simply becomes \be L = \sqrt{-g} \left[\frac{1}{4} g^{\mu
\nu} g^{\alpha \rho} g^{\beta \sigma} \nabla_\mu h_{\alpha \beta}
\nabla_\nu h_{\rho \sigma} - \frac{1}{2} h_{\alpha \beta} h_{\mu
\nu} R^{\beta \mu \alpha \nu} + \frac{1}{2} g^{\mu \nu} \partial_\mu
\phi
\partial_\nu \phi \right]. \ee We will henceforth drop the (re-defined) dilaton
field, since it decouples from the amplitudes we are interested in.

Mirroring the photon-graviton analysis above, we will now introduce
a vielbein for the background field.  In order to make a clear
distinction between `left' and `right' indices, we will use a `left'
vielbein $e$ and a `right' vielbein $\tilde{e}$, which introduces
two copies of Lorentz gauge redundancies with their associated
connections $\omega, \tilde{\omega}$. We then write \be h_{\mu \nu}
= e^a_\mu \tilde{e}^{\tilde{a}}_\nu h_{a \tilde{a}} , \
\nabla_\alpha h_{\mu \nu} = e^a_\mu \tilde{e}^{\tilde{a}}_\nu
D_\alpha h_{a \tilde{a}} \ee with \be D_\alpha h_{a \tilde{a}} =
\partial_\alpha h_{a \tilde{a}} + {\omega^b_{\alpha a}} h_{b
\tilde{a}} + {\tilde{\omega}^{\tilde{b}}_{\alpha \tilde{a}}}h_{a
\tilde{b}} . \ee Of course in reality $e = \tilde{e}$ and $\omega =
\tilde{\omega}$, but this simple notation will help to keep track of
the fact that left-right index contractions never occur.  The
lagrangian becomes \be L = \sqrt{-g} \left[\frac{1}{4} g^{\mu \nu}
\eta^{ab} \tilde{\eta}^{\tilde{a} \tilde{b}} D_\mu h_{a \tilde{a}}
D_\nu h_{b \tilde{b}} - \frac{1}{2} h_{a \tilde{a}} h_{b \tilde{b}}
R^{a b \tilde{a} \tilde{b}}\right]. \ee As in the photon-graviton
case above, we choose light-cone gauge so that \be \omega^+_{a b} =
\tilde{\omega}^+_{\tilde{a} \tilde{b}} = g^{++} = g^{+i} = 0 \ \ \
\mathrm{and} \ \ \ g^{+-} = 1 , \ee and there are no $O(z^2)$
vertices and the only $O(z)$ vertices preserve both the spin Lorentz
symmetries -- except for the unique set of diagrams that give
contributions up to $O(z^2)$. The $O(z^2)$ terms must come from the
two derivative part of the lagrangian, which don't break either of
the left or right spin ``Lorentz" invariances, and are thus
proportional to $\eta^{ab} \tilde{\eta}^{\tilde{a} \tilde{b}}$. The
$O(z)$ terms that violate the symmetry come from a derivative on $h$
and a single $\omega$ or $\tilde{\omega}$ insertions, and hence have
the form $\eta^{ab} \tilde{A}^{\tilde{a} \tilde{b}} + A^{a b}
\tilde{\eta}^{\tilde{a} \tilde{b}}$ where $A,\tilde{A}$ are
antisymmetric. Now consider the $O(1)$ parts of the amplitude. Since
all propagators scale as $1/z$, these can only come directly from
the $O(1)$ vertices in the Lagrangian. There are terms involving
$\omega^2$ or $\tilde{\omega}^2$, each of which breaks one of the
``Lorentz" symmetries but not the other, so these give amplitudes of
the form $\eta^{ab} B^{\tilde{a} \tilde{b}} + B^{a b}
\tilde{\eta}^{\tilde{a} \tilde{b}}$. There are also terms
proportional to $\omega \tilde{\omega}$ and the Riemman tensor,
which are antisymmetric separately in $(ab)$ and $(\tilde{a}
\tilde{b})$, and which thus give a contribution to the amplitude
$A^{a b \tilde{a} \tilde{b}}$ which has the same antisymmetry
properties. Thus we find \be M^{a \tilde{a} b \tilde{b}} = c z^2
\eta^{ab} \tilde{\eta}^{\tilde{a} \tilde{b}} + z \left(\eta^{ab}
\tilde{A}^{\tilde{a} \tilde{b}} + A^{ab} \tilde{\eta}^{\tilde{a}
\tilde{b}} \right) + A^{a b \tilde{a} \tilde{b}} + \eta^{ab}
\tilde{B}^{\tilde{a} \tilde{b}} + B^{ab} \tilde{\eta}^{\tilde{a}
\tilde{b}} + \frac{1}{z} C^{a b \tilde{a} \tilde{b}} + \cdots \ee
where $A^{ab}$ is an antisymmetric matrix, $B^{ab}$ is an arbitrary
matrix, and $A^{a b \tilde{a} \tilde{b}}$ is antisymmetric in $(ab)$
and $(\tilde{a} \tilde{b})$. It is quite remarkable that this
symmetry structure is precisely what we would get from `squaring'
the Yang-Mills ansatz above -- for instance the $\eta^{ac}
\tilde{B}^{\tilde{b} \tilde{d}}$ type terms come from multiplying
the $z \eta^{ab}$ pieces in $M^{ab}_{\mathrm{gauge}}$ with the
$(1/z) B^{\tilde{a} \tilde{b}}$ terms in $M^{\tilde{a}
\tilde{b}}_{\mathrm{gauge}}$, while the symmetry structure of $A^{a
b \tilde{a} \tilde{b}}$ arises from the product of the two
anti-symmetric matrices, $A^{ab} \tilde{A}^{\tilde{a} \tilde{b}}$.

Having established this ansatz for $M^{a b \tilde{a} \tilde{b}}$, we
contract with the graviton polarization tensors to obtain the
physical amplitudes. We again use the Ward identity to further
simplify the amplitude.  The identity says that \be p_{j a}(z) M^{a
\tilde{a}, b \tilde{b}} \epsilon_{k b \tilde{b}} = 0 , \ee so as in
the gauge case, we can use this to show that \be (p_{j a} + z q_a)
M^{a \tilde{a}, b \tilde{b}} \epsilon_{k b \tilde{b}} = 0
 \ \ \ \implies \ \ \ q_a M^{a \tilde{a}, b \tilde{b}} \epsilon_{k b \tilde{b}}
 = -\frac{1}{z} p_{j a} M^{a \tilde{a}, b \tilde{b}} \epsilon_{k b \tilde{b}}.
\ee This means that we can take \be \epsilon_{j a \tilde{a}}^{--} =
q_a q_{\tilde{a}} \to \frac{1}{z^2} p_{j a} p_{j \tilde{a}} \ee when
we compute $(--, h)$ amplitudes.  A quite non-trivial example is \be
M^{--,--}(z) &=& \epsilon_{j a \tilde{a}}^{--} M^{a \tilde{a}, b
\tilde{b}} \epsilon_{k b \tilde{b}}^{--} = \frac{1}{z^2} p_{j a}
p_{j \tilde{a}} M^{a \tilde{a}, b \tilde{b}}
(q_b^* + z p_{j b}) (q_{\tilde{b}}^* + z p_{j \tilde{b}}) \nonumber \\
&=& \frac{1}{z} C^{a b \tilde{a} \tilde{b}} p_{j a} p_{j \tilde{a}}
p_{j b} p_{j \tilde{b}} + O(\frac{1}{z^2}) \to \frac{1}{z} \ee as $z
\to \infty$. This is good enough to obtain recursion relations,
though a little extra work shows that $C^{a b \tilde{a} \tilde{b}}$
is not a completely generic tensor but is a sum of terms
antisymmetric in $(ab)$, and in $(\tilde{a} \tilde{b})$, so that
even the $O(1/z)$ term above vanishes and the leading amplitude
scales as $1/z^2$. Other results are similar and, as we noted above,
they conform to the pattern $M_{{\mathrm grav}} \sim M_{{\mathrm
gauge}} \times M_{{\mathrm gauge}}$. For general two-index
polarization tensors, giving amplitudes for gravitons as well as
dilatons and antisymmetric tensor fields, we find the scaling
\begin{center}
\begin{tabular}{|c|c|c|c|c|c|c|}
\hline
$\epsilon_1 \backslash \epsilon_2$&$- -$&$- +$&$++$&$-$T&$+$T&TT\\
\hline \hline
$- -$&$1/z^2$&$1/z^2$&$1/z^2$&$1/z^2$&$1/z^2$&$1/z^2$\\
\hline
$- +$&$z^2$&$z^2$&$1/z^2$&$z^2$&$1$&$1$\\
\hline
$+ +$&$z^6$&$z^2$&$1/z^2$&$z^4$&$1$&$z^2$\\
\hline
$-$T&$1$&$1$&$1/z^2$&$1$&$1$ or $1/z$&$1$ or $1/z$\\
\hline
$+$T&$z^4$&$z^2$&$1$&$z^4$ or $z^3$&$1$&$z^2$ or $z$\\
\hline
TT&$z^2$&$1$&$1/z^2$&$z^2$ or $z$&$1$ or $1/z$&$z^2$, $z$, or $1$\\
\hline
\end{tabular}
\end{center}
where the various possibilities involving $T$ polarizations depend
on whether or not the $T$ factors in the graviton polarization
tensors are parallel or orthogonal. We have checked that these
accord with behavior of the $2 \to 2$ gravitational scattering
amplitudes in arbitrary $D$. Since $M(z \to \infty)$ vanishes for
all $(--, h)$ helicity combinations, the BCFW Recursion Relations
hold in general relativity for all dimensions $D \ge 4$.

\section{Discussion}
We close with brief comments on some possible implications of our
results. The ability to use BCFW recursion relations to compute
higher-dimensional amplitudes can be useful for computing certain
massive 4D amplitudes where the massive particles can be thought of
as KK modes in the dimensional reduction of the higher-dimensional
theory (other extensions of recursion relations to include massive
particles have been discussed in \cite{mass1,mass2}). This can be
used for the analytic computation of some massive SM amplitudes of
relevance to the LHC. For instance, to compute $g g \to t \bar{t} +
n g$, we can consider a 5D amplitude with all massless particles
where all the gluon momenta are four-dimensional but the 5D top
quarks carry five-momentum. However, while it is nice to have
analytic expressions for these amplitudes, it does not really
appreciably help with bread and butter QCD physics, as the
amplitudes can be numerically determined in any case. The real
bottleneck is not in determining amplitudes at tree level, but in
performing the phase space integrals needed to convert the
amplitudes to rates. Nevertheless, our discussion does suggest
interesting avenues for further theoretical exploration.

Our analysis of the large $z$ scaling of tree amplitudes relied
heavily on the form of the Yang-Mills and Gravity Lagrangians.
However we know that the structure of Yang-Mills and Gravity are
forced on us by consistent S-matrices for massless spin 1 and spin 2
particles (with minimal derivative interactions). It must therefore
be possible and illuminating to determine the large $z$ scalings
directly from S-matrix arguments, without passing through the
Lagrangian as an intermediary.

While the BCFW recursion relations beautifully realize the S-matrix
program for tree amplitudes, the situation at loop level is not
quite as transparent even though much is understood \cite{loop}. One
issue, for instance, is the analytic structure of $M(z)$, where in
addition to expected poles and cuts, there are also `unreal' poles
without a clear physical interpretation. It would be interesting to
see if our picture sheds any further light on this. Furthermore, the
large $z$ scaling of amplitudes is modified at loop level but, in
examples, continues to exhibit very interesting patterns that would
be interesting to understand along the same lines as our tree-level
analysis.

A related issue is the much better than expected behavior that has
been found for gravity amplitudes at loop level. The most intriguing
recent example of this has been for $N=8$ supergravity, where
cancelations not obviously guaranteed by SUSY have led some to
conjecture that the theory might even be perturbatively finite
\cite{Bern1,Bern2}. But even pure gravity amplitudes appear to be
better behaved than expected by naive power-counting \cite{Bern3}, \cite{BjerrumBohr2}.
This seems very surprising, especially since in the usual view,
power-counting is controlled purely by Wilsonian dimensional
analysis, and does not care about whether we have a ``simple"
non-renormalizable theory of scalars like the chiral Lagrangian, or
a more ``complicated" one such as gravity. However it appears that
precisely these ``complicated" theories might have unexpectedly good
UV behavior! Why should this be?

A possible clue is that these cancelations have progenitors in the
soft large $z$ behavior of tree amplitudes we have discussed in this
paper \cite{Bern3}, which arise as cuts of loop diagrams. But we
have understood why certain graviton amplitudes are softer at
infinite (complex) momentum than gauge amplitudes which are in turn
softer than the scalar ones -- at infinite momentum, the amplitudes
for spin $s$ particles are governed by $s$ copies of spin ``Lorentz"
symmetries -- so the more ``complicated" theories with higher spin
are actually constrained by larger symmetries. It is tempting to
speculate that these enhanced symmetries are further extended in
theories with high degrees of supersymmetry and can help illuminate
the mysterious cancelations found in $N=8$ supergravity.

We are grateful to Zvi Bern and Michael Peskin for triggering our
interest in this subject. We also thank Christian Bauer, Niels 
Bjerrum-Bohr, Richard Brower, Michelangelo Mangano, Iain Stewart, 
Edward Witten and especially
Freddy Cachazo and  Juan Maldacena for stimulating discussions. We
further thank Zvi Bern and Lance Dixon both for enlightening
correspondence as well as for very helpful comments on the draft.
The work of N.A.-H. is supported by the DOE under grant
DE-FG02-91ER40654, and J.K. is supported by a Hertz foundation
fellowship and an NSF fellowship.

\end{document}